\documentclass[aps,pre,onecolumn]{revtex4-1}
\usepackage{graphicx}
\usepackage{dcolumn}
\usepackage{bm}
\usepackage{amsmath}
\usepackage{mathrsfs}
\usepackage[dvips]{color}

\begin{document}

\title{Disassortativity of percolating clusters in random networks}
\author{Shogo Mizutaka}\email{shogo.mizutaka.sci@vc.ibaraki.ac.jp}\author{Takehisa Hasegawa}\email{takehisa.hasegawa.sci@vc.ibaraki.ac.jp}
\affiliation{Department of Mathematics and Informatics, Ibaraki University, 2-1-1 Bunkyo, Mito 310-8512, Japan}

\date{\today}
\begin{abstract}
We provide arguments for the property of the degree-degree correlations of giant components formed by the percolation process on uncorrelated random networks. 
Using the generating functions, we derive a general expression for the assortativity of a giant component, $r$, which is defined as Pearson's correlation coefficient for degrees of directly connected nodes.
For uncorrelated random networks in which the third moment for the degree distribution is finite, we prove the following two points. 
(1) Assortativity $r$ satisfies the relation $r\le 0$ for $p\ge p_{\rm c}$. 
(2) The average degree of nodes adjacent to degree $k$ nodes at the percolation threshold is proportional to $k^{-1}$ independently of the degree distribution function.
These results claim that disassortativity emerges in giant components near the percolation threshold. 
The accuracy of the analytical treatment is confirmed by extensive Monte Carlo simulations.
\end{abstract}
\maketitle

\section{Introduction}

All systems are considered as networks if they consist of elements, and the relation between the elements can be defined.
Owing to the generality of the definition of networks, various systems such as ecosystems, metabolic interactions, the World Wide Web, and social relationships are regarded as networks. 
Thus far, network science has extracted common properties from real networks \cite{Albert02,Dorogovtsev02}.
A representative one is the correlation between degrees of directly connected nodes \cite{Newman02,Newman03}.
If similar (dissimilar) degree nodes are more likely to connect to each other in a network, the network has positive (negative) degree-degree correlation. 
We often call a network with positive (negative) degree-degree correlation an assortative (disassortative) network. 
Newman discovered that social networks possess positive degree correlations whereas biological and technological networks are disassortative \cite{Newman02}. 
Following the seminal work of Newman, the degree correlations of complex networks have been studied extensively.
One of the reasons for this is that the degree correlations affect the behavior of dynamics on networks. Much effort has been devoted to examining the relation between the degree-degree correlation and phenomenological models on networks such as failures, spreading of diseases or information, and synchronization, to gain a deep understanding of the character of real-world networks \cite{Gleeson08,Goltsev08,Arenas08,Satorras15,Melnik14}.

There are networks in which no direct path along edges exists between two nodes. Such networks consist of several connected components.  
It is noticed that the degree correlation of a component is different from that of the whole network if the network is not singly connected.
Recent works have formalized the joint probability of degrees in the giant component (GC) whose size is proportional to that of the whole network by the generating function method and obtained the average degree $\bar{k}_{\rm nn}(k)$ of nodes adjacent to degree $k$ nodes \cite{Bialas08} and the assortativity $r$ defined by Pearson's correlation coefficient for nearest degrees \cite{Tishby18}. 
As  demonstrated for some random networks \cite{Tishby18,Bialas08}, the GC can have the negative degree-degree correlation (\textit{disassortativity}) in spite that the whole network is degree-uncorrelated.
In addition, Tishby \textit{et al.} have shown that the correlation between degrees for the GC in the Erd\H{o}s-R\'enyi random graph is always negative if the network is not singly connected and the average degree is greater than unity or, equivalently, the GC exists \cite{Tishby18}.

The above generating function method can be generalized to the case of the percolation problem {\it on} given substrate networks.
In the percolation problem on networks, each node is occupied (not removed) with a given probability and is unoccupied (removed) otherwise. 
It is known that the system undergoes the emergence of a percolating cluster, i.e., a GC of occupied nodes, at a certain value of occupation probability called as the percolation threshold.
It is, however, unknown what correlation percolating clusters on uncorrelated random networks exhibit, especially at and around the percolation threshold where the system exhibits critical behavior \cite{Stauffer}.

In this study, we analyze the degree correlation of the GC generated by the site percolation process on uncorrelated random networks with arbitrary degree distribution.
It is already known that the site percolation process on uncorrelated networks does not induce any degree-degree correlation as long as we focus on the degree-degree correlation of the whole network consisting of occupied nodes \cite{Srivastava12}. We extract the GC from the whole network and examine what degree-degree correlation is observed from the GC.
By formulating the generating function for the joint probability of degrees of the GC, we prove that the GC in random networks with arbitrary degree distribution $P(k)$ always shows disassortativity in terms of assortativity $r$ if the third moment $\langle k^3\rangle$ of $P(k)$ is finite and the networks are not singly connected. 
In addition, by analyzing the average degree $\bar{k}_{\rm nn}(k)$ of nodes adjacent to degree $k$ nodes, we show that $\bar{k}_{\rm nn}(k)$ at the percolation threshold is proportional to $k^{-1}$ as long as $\langle k^3\rangle<\infty$ and is also a decreasing function of $k$ near $p=p_{\rm c}$.
These results mean that the GC possesses disassortativity near the percolation threshold.
The validity of the analytical treatment is confirmed by extensive Monte Carlo simulations.

The rest of this paper is organized as follows. 
In Sec.~\ref{sec:analytical}, we formulate the assortativity $r$ for the GC created by site percolation using the generating functions. 
The comparison between analytical treatment proposed in Sec.~\ref{sec:analytical} and simulations is shown in Sec.~\ref{sec:Numerical}. 
In addition, we show exact expressions of assortativity $r$ at the critical point for $z$-regular random networks and Erd\H{o}s-R\'{e}nyi random graphs. 
In Sec.~\ref{sec:knn}, we further show the disassortativity of the GC by showing that $\bar{k}_{\rm nn}(k)$ is a decreasing function of degree $k$. 
Section \ref{sec:Summary} is devoted to the summary and discussion.

\section{Analytical treatments \label{sec:analytical}}

Let us consider an uncorrelated random network with an arbitrary degree distribution $P(k)$. 
First, let $G_{0}(x)$ be the generating function for the probability, $P(k)$, of a randomly chosen node having degree $k$, as 
\begin{equation}
	G_{0}(x)=\sum_{k=0}^{\infty}P(k)x^k.
	\label{eq:g0}
\end{equation}
Using Eq.~(\ref{eq:g0}), the generating function $G_{1}(x)$ for the probability of an edge leading to a degree $k$ node is given by
\begin{eqnarray}
	G_{1}(x) &=&G'_{0}(x)/G'_{0}(1) \nonumber \\
	&=&\sum_{k=1}^{\infty}\frac{kP(k)}{\langle k\rangle}x^{k-1},
	\label{eq:g1}
\end{eqnarray}
where $G'_{0}(x)$ is the derivative of $G_{0}(x)$ with respect to $x$ and $\langle k\rangle$ is the mean of the degree distribution $P(k)$, $\langle k \rangle=\sum_k k P(k)$. 
In this study, we concentrate on the site percolation problem on a given substrate network with $P(k)$: each node is occupied with probability $p$ and is unoccupied otherwise.
In general, there exists a threshold $p_{\rm c}$ above which an infinitely large cluster, i.e., a GC, emerges in the thermodynamic limit, which means that the fraction $S$ of nodes belonging to the GC becomes $S>0$ from $p> p_{\rm c}$.
We denote by $u$ the probability that one end of an edge randomly chosen from the substrate network does not lead to a GC. 
The probability $u$ is given as the solution of the following self-consistent equation:
\begin{equation}
	u=q+pG_{1}(u)
	\label{eq:u}
\end{equation}
where $q=1-p$. Using the probability $u$, we have the fraction $S$ as
\begin{equation}
	S=p\left(1-G_{0}(u)\right).
	\label{eq:S}
\end{equation}
The percolation threshold $p_{\rm c}$ is given with the condition that Eq.~(\ref{eq:u}) has a nontrivial solution of $u<1$, yielding $S>0$. For uncorrelated random networks, it is known as $p_{\rm c}=\langle k \rangle/\langle k(k-1) \rangle$ (see Refs.~\cite{Cohen00,Newman01}).

Let us focus on only degree correlations of GCs formed by the site percolation in uncorrelated networks.
First, we consider the conditional probability $P({\rm GC},k,k'|l,m)$ that a randomly chosen edge has two ends with degree $k$ and $k'$ and belongs to the GC conditioned on the two ends having originally $l$ and $m$ neighbors in a substrate network. 
As $pG_{1}(u)$ is the probability that one end of an edge is occupied and does not lead to the GC, $p^2(1-G_{1}^{k-1}(u)G_{1}^{k'-1}(u))$ represents the probability that a randomly chosen edge leads to two occupied ends with degree $k$ and $k'$ and belongs to the GC.
Therefore, we can write the probability $P({\rm GC},k,k'|l,m)$ as
\begin{widetext}
\begin{equation}
	P({\rm GC},k,k'|l,m)=p^2(1-G_{1}^{k-1}(u)G_{1}^{k'-1}(u))\binom{l-1}{k-1}p^{k-1}q^{l-k}\binom{m-1}{k'-1}p^{k'-1}q^{m-k'}.
\end{equation}	
\end{widetext}
Let $P(k,k')$ and $P(\rm GC)$ be the joint distribution of degrees in the substrate network and the probability that an edge belongs to the GC, respectively. 
The relations $P({\rm GC},k,k')=\sum_{l,m}P({\rm GC},k,k'|l,m)P(l,m)$ and $P(k,k'|{\rm GC})=P({\rm GC},k,k')/P({\rm GC})$ are satisfied. 
We also have $P({\rm GC})=p^2(1-G_{1}^{2}(u))$ immediately. 
For convenience, we denote $P(k,k'|{\rm GC})$ as $P_{\rm GC}(k,k')$ and the subscript GC is used for conditional probabilities conditioned on the GC. 
Using these relations, we find the joint distribution $P_{\rm GC}(k,k')$ of degrees on the GC as
\begin{widetext}
\begin{equation}
	P_{\rm GC}(k,k') = \frac{1-G_{1}^{k-1}(u)G_{1}^{k'-1}(u)}{1-G_{1}^{2}(u)}
	\sum_{l\ge k,~m\ge k'}\binom{l-1}{k-1}p^{k-1}q^{l-k}\binom{m-1}{k'-1}
	p^{k'-1}q^{m-k'}\frac{lP(l)}{\langle k\rangle}\frac{mP(m)}{\langle k\rangle},
	\label{eq:pkk'_gc}
\end{equation}
where we use the relation $P(k,k')=(kP(k)/\langle k\rangle)(k'P(k')/\langle k\rangle)$ because the substrate network is uncorrelated. 
The generating function $B(x,y)$ for $P_{\rm GC}(k,k')$ is obtained as follows (see the Appendix for details):
\begin{eqnarray}
    B(x,y)&=& \sum_{k \ge 1,k' \ge 1} P_{\rm GC}(k,k')x^{k-1}y^{k'-1} \nonumber \\
	&=&\frac{G_{1}(q+px)G_{1}(q+py)-
	G_{1}(q+pG_{1}(u)x)G_{1}(q+pG_{1}(u)y)}{1-G_{1}^{2}(u)}.
	\label{eq:Bxy}
\end{eqnarray}	
\end{widetext}
From $B(x,y)$, the generating function $S(x) [=B(x,1)=B(1,x)]$ for the marginal distribution $Q_{\rm GC}(k) [=\sum_{k'}P_{\rm GC}(k,k')]$, which is the probability of an edge reaching a node with degree $k$ conditioned on the edge in the GC, is
\begin{eqnarray}
	S(x) &=&\sum_{k=1}^{\infty}Q_{\rm GC}(k)x^{k-1} \nonumber \\
	&=&\frac{G_{1}(q+px)-G_{1}(u)G_{1}(q+pG_{1}(u)x)}{1-G_{1}^{2}(u)}.
	\label{eq:Sx}		
\end{eqnarray}
Obviously, these generating functions $B(x,y)$ and $S(x)$ are reduced to expressions for generating functions in Ref.~\cite{Tishby18} when $p=1$.
Thus, the present formalism is a generalization of the previous method, in which the site percolation process is incorporated.
In accordance with the argument in Ref.~\cite{Tishby18}, assortativity $r$ of the GC is given by $B(x,y)$ and $S(x)$ as
\begin{equation}
	r=\frac{\partial_{x}\partial_{y}B(x,y)-\left[ \partial_{x}S(x) \right]^2}
	{\left(x\partial_{x}\right)^2S(x)-\left[ \partial_{x}S(x) \right]^2}\Bigg|_{x=y=1}
	\label{eq:assortativity}.
\end{equation}
Substituting Eqs.~(\ref{eq:Bxy}) and (\ref{eq:Sx}) into Eq.~(\ref{eq:assortativity}), we can find the general result for assortativity of the GC as
\begin{widetext}
\begin{equation}
	r=\frac{-p \tilde{g}_{0}^{2}(g_{1}-\tilde{g}_{1})^2}{(1-\tilde{g}_{0}^{2})
	(g_{1}-\tilde{g}_{0}^{2}\tilde{g}_{1}+pg_{2}-pg^{3}_{0}\tilde{g}_{2})-p(g_{1}-\tilde{g}_{0}^{2}\tilde{g}_{1})^2},
	\label{eq:r2}
\end{equation}
\end{widetext}
where 
\begin{equation}
	g_{n}=G_{1}^{(n)}(1)=\frac{\partial^{n}G_{1}(x)}{\partial x^{n}} \Big|_{x=1}
\end{equation}
and
\begin{equation}
	\tilde{g}_{n}=G_{1}^{(n)}(u)=\frac{\partial^{n}G_{1}(x)}{\partial x^{n}} \Big|_{x=u}.
\end{equation}
The denominator of the right-hand side in Eq.~(\ref{eq:r2}) is equal to $(1-\tilde{g}_{0}^{2})\sigma_{Q_{\rm GC}}/p$, where $\sigma_{Q_{\rm GC}}$ is the variance of $Q_{\rm GC}(k)$, and is a positive real number.
Then, the sign of assortativity is determined by the numerator. 
Therefore, the assortativity satisfies an inequality $r\le 0$ for $p_{\rm c} \lesssim p\le 1$. 
The factor $(g_{1}-\tilde{g}_{1})^2$ in Eq.~(\ref{eq:r2}) claims that if a GC exists, it always exhibits disassortativity independently of the degree distribution because $(g_{1}-\tilde{g}_{1})^2$ becomes a non-zero positive value for $p \gtrsim p_{\rm c}$.
The result is persistent even at $p=1$ when the substrate network is not singly connected, which is consistent with previous results in Refs.~\cite{Tishby18,Bialas08}.
The zero assortativity is observed only when the network is singly connected at $p=1$ because then the factor $\tilde{g}_{0}$ becomes zero.
It is noted here that assortativity $r$ cannot be negative in infinitely large networks with $\langle k^3 \rangle=\infty$ (see Ref.~\cite{Litvak13}). 
The factor $g_{2}$ appearing in the denominator contains $\langle k^3\rangle$ and reflects the feature.

\section{Numerical check \label{sec:Numerical}}
To evaluate the validity of our analytical treatment for uncorrelated random networks, we compare analytical estimates of the assortativity $r$ with corresponding simulation results. 
In our simulations, we utilize the configuration model which realizes uncorrelated random networks according to a predefined degree distribution.
In the following subsections, we concentrate on typical examples, i.e., $z$-regular random graphs, Erd\H{o}s-R\'{e}nyi random graphs, and scale-free networks.
\begin{figure}[t]
\begin{center}
\includegraphics[width=.45\textwidth]{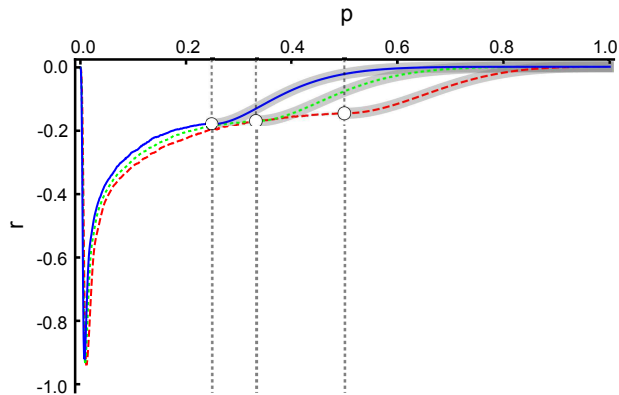}
\caption{Comparison with the analytical treatment and simulation results for $p$ dependence of assortativity $r$. 
The $z$-regular random graph is utilized as the substrate network. 
The grayscale tube lines represent the results obtained by Eq.~(\ref{eq:r2}) with the aid of Eqs.~(\ref{eq:g1}) and (\ref{eq:u}). 
The simulation results are for $z$-regular random graphs with $z=5$ (solid blue line), $4$ (dotted green line), and $3$ (dashed red line). The number of nodes used for simulations is $N=10^6$.
}
\label{fig:zrrg}
\end{center}
\end{figure}

\subsection{$z$-regular random graphs}

First, let us consider $z$-regular random graphs as a simple illustrative example.
The degree distribution $P(k)$ of the $z$-regular random graph is
\begin{equation}
	P(k)=\delta_{kz},
	\label{eq:zrrg}
\end{equation}
whose percolation threshold is given as
\begin{equation}
	p_{\rm c}=\frac{1}{z-1}.
	\label{eq:zrrg_pc}
\end{equation}

Figure~\ref{fig:zrrg} shows the $p$ dependence of assortativity $r$. 
Grayscale tube lines represent analytical estimates obtained from Eq.~(\ref{eq:r2}), and the other lines are drawn from Monte Carlo simulations.
In our simulations, we generated 10 network realizations and performed site percolation $10^3$ times on each realization 
to take the average of $r$ at given values of $p$.
On each run, we specify the largest component, which corresponds to the GC for $p > p_{\rm c}$, based on the Newman-Ziff algorithm \cite{Newman01-2}. The assortativity of the largest component is evaluated and compared with the result obtained by analytical treatment.
Our analytical estimates for $r$ match perfectly with the numerical data for $p > p_{\rm c}$ in all cases.
The vertical dashed lines from left to right indicate the percolation thresholds $p_{\rm c}$ when $z=5$, $4$, and $3$, respectively. 
Our numerical data assert that the assortativity $r$ does not show the singular behavior just at and around the percolation threshold $p_{\rm c}$ even when the system size goes to infinitely large. 
This implies that the analytical expression for the assortativity $r$ at $p=p_{\rm c}$ can be obtained. 
Approximating the probability $u$ at $p\gtrsim p_{\rm c}$ as $u \sim 1-\epsilon$ where $\epsilon$ is an infinitesimal value, we have the relation
\begin{equation}
G_{1}(u)\sim 1-\frac{\langle k(k-1)\rangle}{\langle k\rangle}\epsilon.
\label{eq:g1u}	
\end{equation}
The assortativity $r_{\rm c}$ at $p=p_{\rm c}$ is given by substituting Eq.~(\ref{eq:g1u}) into Eq.~(\ref{eq:r2}) and taking $\epsilon \to 0$ as
\begin{widetext}
\begin{equation}
	r_{\rm c}=-\frac{p_{\rm c}g_{2}^{2}}{2g_{1}(2g_{1}^{2}+g_{2}+
	3p_{\rm c}g_{1}g_{2}+p_{\rm c}g_{3})-p_{\rm c}(2g_{1}^{2}+g_{2})^2}.
	\label{eq:ratpc}
\end{equation}
\end{widetext}
Using Eqs.~(\ref{eq:zrrg}), (\ref{eq:zrrg_pc}), and (\ref{eq:ratpc}), we have the assortativity $r_{\rm c}$ at $p=p_{\rm c}$ for a $z$-regular random graph,
\begin{equation}
	\label{eq:zrrg_rc}
	r_{\rm c}=-\frac{z-2}{5z-8}.
\end{equation}
Large symbols on the edges of grayscale tube lines in Fig.~\ref{fig:zrrg} are $r_{\rm c}$ given by Eq.~(\ref{eq:zrrg_rc}) to confirm the accuracy of analytical treatment.

\begin{figure}[t]
\begin{center}
\includegraphics[width=.45\textwidth]{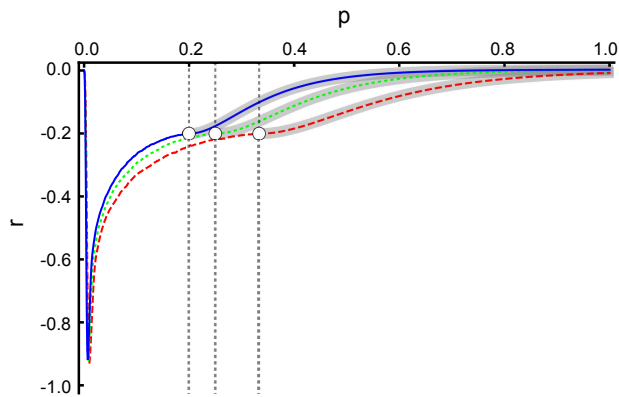}
\caption{Comparison with the analytical treatment and simulation results for $p$ dependence of assortativity $r$.
The substrate network obeys the degree distribution with Eq.~(\ref{eq:ERRG}). 
The grayscale tube lines represent the results obtained by Eq.~(\ref{eq:r2}) with the aid of Eqs.~(\ref{eq:g1}) and (\ref{eq:u}).
The simulation results are for Erd\H{o}s-R\'{e}nyi random graphs with $\langle k\rangle=5$ (solid blue line), $4$ (dotted green line), and $3$ (dashed red line).
The number of nodes used for simulations is $N=10^6$.
Ten realizations of networks were generated, and the site percolation process was performed $10^3$ times for each network to obtain simulation results.
}
\label{fig:ERgraphs}
\end{center}
\end{figure}
\subsection{Erd\H{o}s-R\'{e}nyi random graphs}

The degree distribution $P(k)$ and the percolation threshold $p_{\rm c}$ for Erd\H{o}s-R\'{e}nyi random graphs are
\begin{equation}
	P(k)=\frac{\langle k\rangle^k e^{-\langle k \rangle}}{k!}
	\label{eq:ERRG}
\end{equation}
and
\begin{equation}
	p_{\rm c}=\frac{1}{\langle k\rangle},
\end{equation}
respectively.
Figure~\ref{fig:ERgraphs} shows the assortativity $r$ as a function of $p$.
The analytical estimates for $r$ match perfectly with the numerical data for $p > p_{\rm c}$ as is the case with $z$-regular random graphs.
The assortativity $r_{\rm c}$ at $p=p_{\rm c}$ for Erd\H{o}s-R\'{e}nyi random graphs is given as
\begin{equation}
	r_{\rm c} =-\frac{1}{5},
\end{equation}
independently of the average degree $\langle k\rangle$ of original graphs.

\begin{figure}[t]
\begin{center}
\includegraphics[width=.45\textwidth]{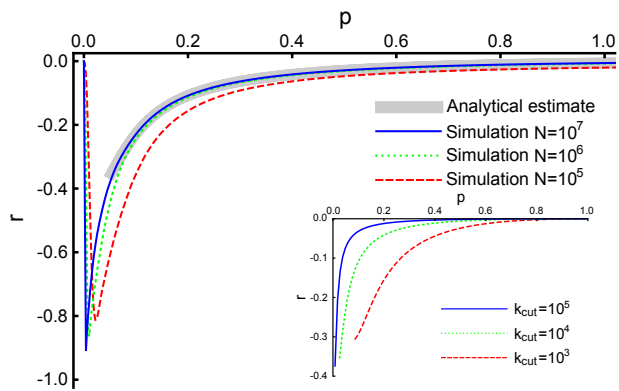}
\caption{Comparison with the analytical treatment and simulation results for $p$ dependence of assortativity $r$. 
Main panel: The results are for scale-free networks with exponent $\gamma=2.5$ and cutoff degree $k_{\rm cut}=10^3$ of the degree distribution.
The grayscale tube line represents the result obtained by Eq.~(\ref{eq:r2}) with the aid of Eqs.~(\ref{eq:g1}) and (\ref{eq:u}). 
The solid blue, dotted green, and dashed red lines are simulation results for $N=10^{7}$, $10^6$, and $10^5$, respectively.
Ten realizations of networks were generated, and the site percolation process was performed $10^3$ times for each network to obtain simulation results.
Inset: The analytical estimates of $r$ for scale-free networks with $\gamma=2.5$ and $k_{\rm cut}=10^5$ (solid blue), $10^4$ (dotted green), and $10^3$ (dashed red).}
\label{fig:SF}
\end{center}
\end{figure}

\subsection{Scale-free networks}

Finally, we consider scale-free networks whose degree distribution obeys $P(k) \sim k^{-\gamma}$ for $2 \le k \le k_{\rm cut}$. 
To argue the effect of network heterogeneity on the degree correlation of the GC, in scale-free networks with $k_{\rm cut} \to \infty$, we start by comparing analytical and numerical results for the case with a finite cutoff degree, i.e., $k_{\rm cut} < \infty$.
Figure~\ref{fig:SF} shows the results for the scale-free networks with exponent $\gamma=2.5$ and cutoff degree $k_{\rm cut}=10^3$. 
Monte Carlo data asymptotically reach the analytical line as increasing the system size $N$, which implies that the analytical treatment is valid for infinite networks with a finite cutoff degree. This also 
indicates the disassortativity of the GCs formed by occupied nodes on the scale-free networks with a finite cutoff degree analytically and numerically.
The validity of our analytical treatment holds for different values of $\gamma$ and $k_{\rm cut}$ (not shown).
Based on the analytical treatment, we display the $p$ dependence of $r$ for scale-free networks with $\gamma=2.5$ and different values of  $k_{\rm cut}$ (the inset of Fig.~\ref{fig:SF}).
It is known for $\gamma<3$ that the percolation threshold approaches zero as $k_{\rm cut}$ increases.
With increasing $k_{\rm cut}$, the trend that the assortativity $r$ goes to $0$ rapidly above $p_{\rm c}$ which is located on the left end of the line, is enhanced.
This result indicates that when $k_{\rm cut}\to \infty$, $p_{\rm c}$ goes to $0$ and the assortativity $r$ becomes $0$ for $p>0$.

\begin{figure}[t]
\begin{center}
\includegraphics[width=.45\textwidth]{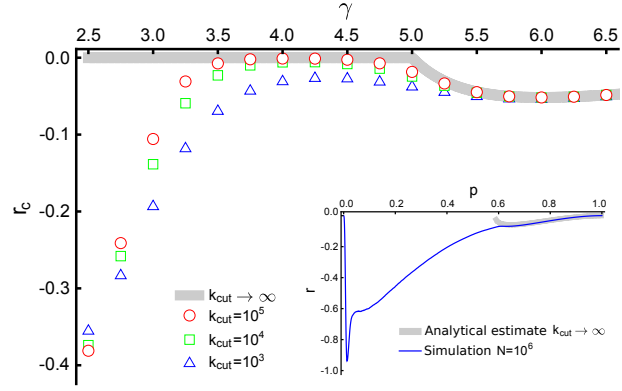}
\caption{Main panel: Analytical estimates of assortativity $r_{\rm c}$ at $p=p_{\rm c}$ as a function of $\gamma$.
The grayscale tube line is the result for scale-free networks with $k_{\rm cut}\to \infty$.
Symbols represent the results for scale-free networks with finite cutoff degree $k_{\rm cut}=10^5$ (red circles), $10^4$ (green squares), and $10^3$ (blue triangles).
All data are obtained by Eq.~(\ref{eq:r2}) with the aid of Eqs.~(\ref{eq:g1}) and (\ref{eq:u}).
Inset: $p$ dependence of the assortativity $r$. The grayscale tube line represents the analytical result for scale-free networks with $\gamma=4.5$ and $k_{\rm cut}\to \infty$ of the degree distribution. The line is the simulation result for $N=10^6$.
Ten realizations of networks were generated, and the site percolation process was performed $10^3$ times for each network to obtain simulation results.
}
\label{fig:r_of_gamma}
\end{center}
\end{figure}
In addition, the assortativity $r_{\rm c}$ at $p=p_{\rm c}$ as a function of the scale-free exponent $\gamma$ is shown in Fig.~\ref{fig:r_of_gamma}.
The grayscale tube line and the symbols represent the analytical estimate of $r_{\rm c}$ for networks without and with finite cutoff degrees, respectively.
For $\gamma>5$,  $p_{\rm c}>0$ and $r_{\rm c}<0$ even for $k_{\rm cut} \to \infty$. Most symbols are on the grayscale tube line, indicating that $r_{\rm c}$ is not sensitive to $k_{\rm cut}$ for $\gamma>5$.
For $3<\gamma<5$, $p_{\rm c}>0$ and $r_{\rm c}=0$ when $k_{\rm cut} \to \infty$.
The fashion that $r_{\rm c}\to 0$ at $k_{\rm cut}\to \infty$ is also reflected on the $k_{\rm cut}$ dependence of $r_{\rm c}$, i.e., the movement of symbols at fixed $\gamma$.
The zero assortativity of the GC is because the right-hand side of Eq.~(\ref{eq:ratpc}) includes $\langle k^4\rangle$, which diverges for $3<\gamma<5$, in the denominator, where $\langle k^4\rangle$ is induced by asymptotically expanding the right-hand side of Eq.~(\ref{eq:r2}) near $p=p_{\rm c}$.
For $\gamma<3$, the $k_{\rm cut}$ dependence of $r_{\rm c}$ in Fig.~\ref{fig:r_of_gamma} seems to suggest that $r_{\rm c}$ converges to a finite negative value as $k_{\rm cut} \to \infty$. However, $p_{\rm c}=0$ in this region and $r_{\rm c}$ will become $0$ for $p>0$, as mentioned above.

Finally, we consider $r$ for $p>p_{\rm c} (>0)$ for the case of $4<\gamma<5$. 
In the inset of Fig.~\ref{fig:r_of_gamma}, we display the $p$ dependence of $r$ for the scale-free network with $\gamma=4.5$.
We find that $r$ always takes a finite negative value at $p>p_{\rm c}$, although the assortativity $r_{\rm c}$ at $p=p_{\rm c}$ becomes $0$.
The assortativity $r$ includes the moments of the degree distribution: 
$\langle k^3 \rangle$ in $r$ of the whole network or its GC for $p>p_{\rm c}$, and $\langle k^4 \rangle$ of the GC at $p=p_{\rm c}$.
Therefore, $r$ sometimes becomes useless for scale-free networks because these moments diverge according to the value of the exponent $\gamma$. 
However, such zero assortativity never means that the GC does not have the degree-degree correlation.
We consider the disassortativity of the GC in scale-free networks with an exponent in $3<\gamma<5$ in the next section.

\section{Behavior of $\bar{k}_{\rm nn}(k)$ \label{sec:knn}}
\begin{figure}[t]
\begin{center}
\includegraphics[width=.45\textwidth]{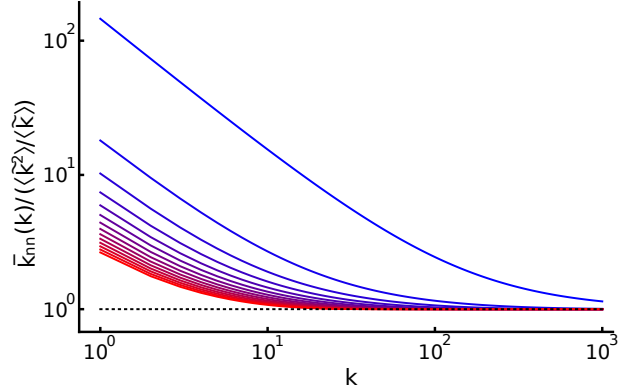}
\caption{Rescaled $\bar{k}_{\rm nn}(k)$ by $\frac{\langle \tilde{k}^2\rangle}{\langle \tilde{k}\rangle}$ as a function of degree $k$ for scale-free networks with $\gamma=3.5$ and $k_{\rm cut} \to \infty$.
Analytical estimates~(\ref{eq:knn}) for $p=0.27, 0.28, \ldots, 0.40$ are shown as the lines from top to bottom. 
Here $p_{\rm c}=\langle k \rangle/\langle k(k-1) \rangle= 0.2687$.
}
\label{fig:knn_of_k}
\end{center}
\end{figure}

We further discuss the disassortativity of the GC with a different quantity. 
The average degree $\bar{k}_{\rm nn}(k)$ of nodes adjacent to degree $k$ nodes is more informative than the assortativity $r$. 
The quantity $\bar{k}_{\rm nn}(k)$ of the GC is calculated from the probability $P_{\rm GC}(k'|k)$ of degree $k'$ nodes adjacent to the degree $k$ nodes in the GC. 
The probability $P_{\rm GC}(k'|k)$ is given by 
\begin{equation}	
	P_{\rm GC}(k'|k)= \frac{1-\tilde{g}_{0}^{k-1}\tilde{g}_{0}^{k'-1}}{1-\tilde{g}_{0}^{k}}\frac{k'\tilde{P}(k')}
	{\langle \tilde{k}\rangle},
	\label{eq:pgck'|k}
\end{equation}	
where $\tilde{P}(k)=\sum_{m\ge k}^{\infty}\binom{m}{k}p^{k}q^{m-k}P(m)$ and $\langle \tilde{k}^n\rangle=\sum k^n\tilde{P}(k)$ (see the Appendix for details).
Note that $\tilde{P}(k)$ corresponds to the degree distribution for the network whose nodes are randomly occupied with probability $p$ on the substrate network.
Equation~(\ref{eq:pgck'|k}) leads to the average degree $\bar{k}_{\rm nn}(k)~[=\sum_{k'}k'P_{\rm GC}(k'|k)]$ of nodes adjacent to degree $k$ nodes as
\begin{equation}
	\bar{k}_{\rm nn}(k)=\frac{\langle \tilde{k}^2\rangle}{\langle \tilde{k}\rangle}\left
	(\frac{1-\tilde{g}_{0}^{k-1}h\left(\tilde{g}_{0}\right)}{1-\tilde{g}_{0}^{k}}\right),
	\label{eq:knn}
\end{equation}
where
\begin{equation}
	h\left(\tilde{g}_{0}\right)=\sum_{k}\frac{k^{2}}{\langle \tilde{k}^2\rangle}\tilde{P}(k)\tilde{g}_{0}^{k-1}.
\end{equation}
Figure~\ref{fig:knn_of_k} shows analytical estimates of rescaled $\bar{k}_{\rm nn}(k)$ by $\langle \tilde{k}^2\rangle/\langle \tilde{k}\rangle$ as the function of degree $k$ for a scale-free network with $\gamma=3.5$ and $k_{\rm cut}\to \infty$ to which the GC shows zero assortativity for $p\ge p_{\rm c}$.
The lines for several values of $p (\gtrsim p_{\rm c})$ indicate that each rescaled $k_{\rm nn}(k)$ decreases monotonically with increasing degree $k$. This means that disassortativity is observed in the GC formed by percolation processes on the scale-free network with $\gamma=3.5$ and $k_{\rm cut}\to \infty$.

Finally, we study the behavior of $\bar{k}_{\rm nn}(k)$ near $p=p_{\rm c}$.
Using Eq.~(\ref{eq:g1u}), we expand Eq.~(\ref{eq:knn}) as follows:
\begin{widetext}
\begin{equation}
	\bar{k}_{\rm nn}(k)\sim \frac{\langle \tilde{k}^2\rangle}{\langle \tilde{k}\rangle}
	\Bigg\{1+\frac{1}{k}\frac{\langle \tilde{k}^3\rangle-2\langle \tilde{k}^2\rangle}
	{\langle \tilde{k}^2\rangle} -\left(1-\frac{1}{k}\right)\frac{\langle \tilde{k}^3\rangle-
	\langle \tilde{k}^2\rangle}{\langle \tilde{k}^2\rangle}\frac{\langle k^2\rangle-\langle k\rangle}
	{\langle k\rangle}\epsilon\Bigg\}.
\end{equation}
\end{widetext}
The result means that $\bar{k}_{\rm nn}(k)$ near $p=p_{\rm c}$ is proportional to $k^{-1}$ independently of the original degree distribution $P(k)$. 
In addition, in the limit of $\epsilon\to 0$, i.e., $p\to p_{\rm c}$, $\bar{k}_{\rm nn}(k)$ is rewritten as
\begin{equation}
	\bar{k}_{\rm nn}(k)=\frac{\langle \tilde{k}^2\rangle}{\langle \tilde{k}\rangle}
	\Bigg\{1+\frac{1}{k}\frac{\langle \tilde{k}^3\rangle-2\langle \tilde{k}^2\rangle}
	{\langle \tilde{k}^{2}\rangle}\Bigg\}.
	\label{eq:knnAtThreshold}
\end{equation}
The exact expression (\ref{eq:knnAtThreshold}) of $\bar{k}_{\rm nn}(k)$ at $p=p_{\rm c}$ holds if $\langle k^3\rangle<\infty$ because $\langle \tilde{k}^3\rangle$ contains $\langle k^3\rangle$.
To summarize, $\bar{k}_{\rm nn}(k)$ at and above $p_{\rm c}$ shows the disassortativity of the GC for scale-free networks with $4<\gamma<5$, although $r_{\rm c}$ failed to capture it. For $3<\gamma<4$, $\bar{k}_{\rm nn}(k)$ is useless just at $p_{\rm c}$ but again shows the disassortativity of the GC above $p_{\rm c}$.

\section{Summary and Discussion \label{sec:Summary}}

In this work, the degree-degree correlations of GCs formed by the site percolation process on uncorrelated random networks have been analyzed. 
By formulating the joint probability of degrees on a GC by means of the generating function, we have shown the following general properties of GCs formed by the percolation process in random networks. 
(1) The assortativity $r$ defined by Pearson's correlation coefficient for degrees satisfies an inequality $r\le 0$ in the percolating phase if the third moment $\langle k^3\rangle$ of the degree distribution is finite. 
(2) The average degree $\bar{k}_{\rm nn}(k)$ of nodes adjacent to degree $k$ nodes at the percolation threshold is proportional to $k^{-1}$ if $\langle k^3\rangle<\infty$.

As has been shown through this work, the negative degree-degree correlation (disassortativity) naturally emerges when we focus on a component of an uncorrelated network. 
It should be noted that one cannot understand the degree-degree correlations of whole networks even if we analyze their components, and one may not be able to understand correctly the behavior of dynamics on networks even if we investigate the dynamics on the components. 
This probably holds true for real networks constructed by data: the difference between the degree correlations of the whole network and of a component would emerge in real-world networks.
It is necessary to pay attention to the lack of data when we analyze real-world networks because a lack of data, expressed as the removals of nodes or edges in percolation processes, would enhance the degree-degree correlations.

The results in this study are consistent with the previous result concerning the relation between fractality and disassortativity of real-world networks \cite{Yook05}. 
The disassortativity of GCs might be established even if an original network has a certain strength of positive degree-degree correlation. However, the behavior of degree-degree correlations of the GCs in assortative networks is not so simple, as will be argued elsewhere~\cite{Hasegawa18}.

We did not discuss the degree-degree correlations of GCs in scale-free networks with $\langle k^3\rangle = \infty$. 
To evaluate the correlations of such networks, Spearman's rank correlation coefficient of degrees has been utilized
\cite{Litvak13,Zhang16,Fujiki17}. 
It is interesting to evaluate degree-degree correlations of GC using Spearman's rank correlation coefficient, although we expect the generality of disassortativity of percolating clusters.

\begin{acknowledgements}
S.M.\ was supported by a Grant-in-Aid for Early-Career Scientists (No.~18K13473) and a Grant-in-Aid for JSPS Research Fellow (No.~18J00527) from the Japan Society for the Promotion of Science (JSPS) for performing this work.
T.H.\ acknowledges financial support from JSPS (Japan) KAKENHI Grant No.~JP16H03939.
\end{acknowledgements}

\appendix*

\section{Derivation of several quantities}

The generating function $B(x,y)$ for $P_{\rm GC}(k,k')$ in Eq.~(\ref{eq:Bxy}) is calculated as follows:
\begin{widetext}
\begin{eqnarray}
	B(x,y)&=& \sum_{k \ge1,k' \ge1}P_{\rm GC}(k,k')x^{k-1}y^{k'-1} \nonumber \\
	&=&\sum_{k \ge1,k' \ge1}\frac{1-G_{1}^{k-1}(u)G_{1}^{k'-1}(u)}{1-G_{1}^{2}(u)}
	\sum_{l\ge k,~m\ge k'}\binom{l-1}{k-1}p^{k-1}q^{l-k}\binom{m-1}{k'-1}
	p^{k'-1}q^{m-k'}\frac{lP(l)}{\langle k\rangle}\frac{mP(m)}{\langle k\rangle}
	x^{k-1}y^{k'-1} \nonumber \\
	&=& \frac{1}{1-G_{1}^{2}(u)}\sum_{l,m}^{\infty,\infty}
	\frac{lP(l)}{\langle k\rangle}\frac{mP(m)}{\langle k\rangle}\sum_{k,k'}^{l,m}
	\left(1-G_{1}^{k-1}(u)G_{1}^{k'-1}(u)\right)\binom{l-1}{k-1}p^{k-1}
	q^{l-k}\binom{m-1}{k'-1}p^{k'-1}q^{m-k'}x^{k-1}y^{k'-1} \nonumber \\
	&=&\frac{1}{1-G_{1}^{2}(u)}\sum_{l,m}^{\infty,\infty}
	\frac{lP(l)}{\langle k\rangle}\frac{mP(m)}{\langle k\rangle}
	\left((q+px)^{l-1}(q+py)^{m-1}-(q+pG_{1}(u)x)^{l-1}
	(q+pG_{1}(u)y)^{m-1}\right) \nonumber \\
	&=&\frac{G_{1}(q+px)G_{1}(q+py)-
	G_{1}(q+pG_{1}(u)x)G_{1}(q+pG_{1}(u)y)}{1-G_{1}^{2}(u)}.
	\label{eq:Bxy2}
\end{eqnarray}
\end{widetext}
Using Eqs.~(\ref{eq:Bxy2}) and (\ref{eq:Sx}), we obtain components constructing the assortativity $r$ as
\begin{widetext}
\begin{eqnarray}
	\partial_{x}\partial_{y} B(x,y)|_{x=y=1}&=&\frac{p^2G'_{1}(q+px)G'_{1}(q+py)-
	p^2G_{1}^{2}(u)G'_{1}(q+pG_{1}(u)x)G'_{1}(q+pG_{1}(u)y)}
	{1-G_{1}^{2}(u)}\bigg|_{x=y=1} \nonumber\\ 
	&=& \frac{p^2 G'^{2}_{1}(1) - p^2G_{1}^{2}(u)G'^{2}_{1}(u)}{1-G_{1}^{2}(u)},
	\label{eq:dxdyB} \\
	\partial_{x}S(x)|_{x=y=1}&=& \frac{pG'_{1}(q+px)-pG_{1}^{2}(u)G'_{1}
	(q+pG_{1}(u)x)}{1-G_{1}^{2}(u)}\bigg|_{x=y=1} \nonumber\\
	&=& \frac{pG'_{1}(1)-pG_{1}^{2}(u)G'_{1}(u)}{1-G_{1}^{2}(u)},
	\label{eq:dxS} \\
	\partial_{x}^{2}S(x)|_{x=y=1}&=& \frac{p^2G''_{1}(q+px)-p^2G_{1}^{3}(u)G''_{1}
	(q+pG_{1}(u)x)}{1-G_{1}^{2}(u)}\bigg|_{x=y=1} \nonumber \\
	&=& \frac{p^2G''_{1}(1)-p^2G_{1}^{3}(u)G''_{1}(u)}{1-G_{1}^{2}(u)},
	\label{eq:dx2S}
\end{eqnarray}
\end{widetext}
where Eq.~(\ref{eq:u}) holds. By substituting Eqs.~(\ref{eq:dxdyB}),
(\ref{eq:dxS}), and (\ref{eq:dx2S}) into Eq.~(\ref{eq:assortativity}), we have Eq.~(\ref{eq:r2}).

The probability $P_{\rm GC}(k'|k)$ is obtained as follows.
As $Q_{\rm GC}(k)=\sum_{k'}P_{\rm GC}(k,k')$ is given by
\begin{equation}
	Q_{\rm GC}(k)=\frac{1-G_{1}^{k}(u)}{1-G_{1}^2(u)}\sum_{l\ge k}\binom{l-1}{k-1}
	p^{k-1}q^{l-k}\frac{lP(l)}{\langle k\rangle},
	\label{eq:qgck}
\end{equation}
we have $P_{\rm GC}(k'|k)=\frac{P_{\rm GC}(k,k')}{Q_{\rm GC}(k)}$ using Eqs.~(\ref{eq:pkk'_gc}) and (\ref{eq:qgck}) as
\begin{widetext}
\begin{eqnarray}	
	P_{\rm GC}(k'|k)&=&\frac{1-G_{1}^{k-1}(u)G_{1}^{k'-1}(u)}{1-G_{1}^{k}(u)}\sum_{m\ge k'}\binom{m-1}{k'-1}
	p^{k'-1}q^{m-k'}\frac{mP(m)}{\langle k\rangle} \nonumber\\
	&=& \frac{1-G_{1}^{k-1}(u)G_{1}^{k'-1}(u)}{1-G_{1}^{k}(u)}\frac{k'}{p\langle k\rangle}\sum_{m\ge k'}\binom{m}{k'}p^{k'}q^{m-k'}P(m) \nonumber\\
	&=& \frac{1-G_{1}^{k-1}(u)G_{1}^{k'-1}(u)}{1-G_{1}^{k}(u)}\frac{k'\tilde{P}(k')}{\langle \tilde{k}\rangle},
	\label{eq:ap:pgck'|k}
\end{eqnarray}	
\end{widetext}
where $\tilde{P}(k)$ is the degree distribution of connected components consisting of occupied nodes in a network with degree distribution $P(k)$,
\begin{equation}
	\tilde{P}(k)=\sum_{m\ge k}\binom{m}{k}p^{k}q^{m-k}P(m).
	\label{eq:Pktilde}
\end{equation}
Then, the first three moments of $\tilde{P}(k)$ are as follows:
\begin{eqnarray}
	\langle \tilde{k} \rangle &=& p\langle k \rangle, \\
	\langle \tilde{k}^2 \rangle &=& p^2\langle k^2 \rangle+pq\langle k \rangle, \\
	\langle \tilde{k}^3 \rangle &=& p^3\langle k^3 \rangle +3p^2q\langle k^2 \rangle+pq(q-p)\langle k \rangle.
\end{eqnarray}
Substituting $\tilde{g}_0=G_1(u)$ into Eq.~(\ref{eq:ap:pgck'|k}), we have Eq.~(\ref{eq:pgck'|k}).

\end{document}